\documentclass[a4paper,UKenglish]{lipics-v2016}

\usepackage{graphicx}
\usepackage{pgfplots}
\usepackage{pgfplotstable}
\usepgfplotslibrary{groupplots}
\pgfplotsset{compat=newest}
\usepackage{caption}
\usepackage[skip=-3pt,belowskip=4pt]{subcaption}
\usepackage{xspace} 
\usepackage{textcomp}
\usepackage{balance}  
\usepackage{xcolor}
\usepackage{amsmath}
\usepackage{algorithm}
\usepackage{algorithmicx}
\usepackage{cite}
\usepackage{float}
\usepackage{tablefootnote}
\usepackage{enumitem} 
\usepackage[noend]{algpseudocode}
\usepackage{url}
\usepackage{hyperref}

\usepackage{siunitx}
\usepackage{amssymb}
\usepackage{color}
\usepackage{booktabs}
\usepackage{xspace}
\usepackage{cleveref}

\usepackage{wrapfig}
\usepackage{times}

\usepackage[compact]{titlesec}
\titlespacing{\section}{0pt}{2ex}{1ex}
\titlespacing{\subsection}{0pt}{1ex}{0ex}
\titlespacing{\subsubsection}{0pt}{0.5ex}{0ex}

\newif\ifcomments        
\commentstrue
\usepackage{our-comments} 

\newcommand{\defn}[1]       {{\textit{\textbf{\boldmath #1}}}}

\newcommand{\rank}{\mbox{\sc rank}}
\newcommand{\select}{\mbox{\sc select}}

\newcommand{\sbu}{PTSelect\xspace}
\newcommand{\cmu}{CS-Poppy\xspace}
\newcommand{\sdsl}{SDSL\xspace}
\newcommand{\cmusbu}{CS-Poppy\_PTSelect\xspace}
\newcommand{\sdslsbu}{SDSL\_PTSelect\xspace}

\title{A Fast x86 Implementation of Select}

\author{Prashant Pandey}
\author{Michael A. Bender}
\author{Rob Johnson}
\affil{Stony Brook University, Stony Brook, NY USA\\
  \texttt{\{ppandey,bender,rob\}@cs.stonybrook.edu}}
\authorrunning{Pandey et al.} 

\Copyright{Prashant Pandey, Michael A. Bender, and Rob Johnson}


\EventEditors{John Q. Open and Joan R. Acces}
\EventNoEds{2}
\EventLongTitle{42nd Conference on Very Important Topics (CVIT 2016)}
\EventShortTitle{CVIT 2016}
\EventAcronym{CVIT}
\EventYear{2016}
\EventDate{December 24--27, 2016}
\EventLocation{Little Whinging, United Kingdom}
\EventLogo{}
\SeriesVolume{42}
\ArticleNo{23}

\begin{document}

\maketitle

\begin{abstract}
  Rank and select are fundamental operations in succinct data
  structures, that is, data structures whose space consumption
  approaches the information-theoretic optimal.  The performance
  of these primitives is central to the overall performance of
  succinct data structures.

  Traditionally, the select operation is the harder to implement
  efficiently, and most prior implementations of select on machine
  words use $50$-$80$ machine instructions. (In contrast, rank on machine
  words can be implemented in only a handful of instructions on
  machines that support \textsc{popcount}.)  However, recently Pandey
  et al.~\cite{PandeyBeJo17} gave a new implementation of machine-word
  select that uses only four x86 machine instructions; two of which
  were introduced in Intel's Haswell CPUs.

  In this paper, we investigate the impact of this new implementation
  of machine-word select on the performance of general
  bit-vector-select.  We first compare Pandey et al.'s machine-word
  select to the state-of-the-art implementations of Zhou et
  al.~\cite{ZhouAnKa13} (which is not specific to Haswell) and Gog et
  al.~\cite{GogPe14} (which uses some Haswell-specific
  instructions). We exhibit a speedup of $2\times$ to $4\times$.

  We then study the impact of plugging Pandey et al.'s machine-word
  select into two state-of-the-art bit-vector-select implementations.
  Both Zhou et al.'s and Gog et al.'s select implementations perform a
  single machine-word select operation for each bit-vector select. We
  replaced the machine-word select with the new implementation and
  compared performance.  Even though there is only a single
  machine-word select operation, we still obtained speedups of $20\%$
  to $68\%$.  We found that the new select not only reduced the number
  of instructions required for each bit-vector select, but also
  improved CPU instruction cache performance and memory-access
  parallelism.
\end{abstract}

\section{ Introduction}

\noindent A \defn{succinct data structure} consumes an amount of space that is
close to the information-theoretically optimal.  More precisely, if $Z$ denotes
the information-theoretically optimal space usage for a given data-structure
specification, then a succinct data structure would use $Z + o(Z)$ space.
There is much research on how to replace traditional data structures
with succinct
alternatives~\cite{RamanRaRa02,CulpepperNaPu10,Sadakane07,Navarro14,
  Elias74,Fano71,GogBeMo14}, especially for memory-intensive
applications, such as genetic databases, newspaper archives,
dictionaries, etc.  Researchers have proposed numerous succinct data
structures, including compressed suffix arrays~\cite{GrossiVi05}, FM
indexes~\cite{FerraginaMa00}, and wavelet
trees~\cite{Navarro14}.

Two basic operations---namely \defn{rank} and
\defn{select}~\cite{Jacobson89}---are commonly used for navigating
within succinct data structures.
For bit vector $B[0,\ldots,n-1]$, \rank$(j)$ returns the number of $1$s in
prefix $B[0,\ldots,j]$ of $B$; \select$(r)$ returns the position of the
$r$th $1$, that is, the smallest index $j$ such that \rank$(j)=r$.
For example, for the 12-bit vector $B[0,\ldots,11]=$\texttt{100101001010},
\rank$(5)=3$, because there are three bits set to one in the 6-bit
prefix $B[0,\ldots,5]$ of $B$, and \select$(4)=8$, because $B[8]$ is the
fourth $1$ in the bit vector.


Researchers have proposed many ways to improve the empirical performance of
rank and select~\cite{GogPe14, GonzalezGrMa05,Vigna08,NavarroPr12,ZhouAnKa13}
because faster implementations of rank and select yield faster succinct data
structures.  Many succinct data structures are asymptotically optimal, but have
poor performance in practice due to the constants involved in bit-vector rank
and select operations~\cite{Vigna08}.
%
%

This paper focuses on a new implementation of select on machine words
and its performance impact on general bit-vector select operations.


\subsubsection*{ Implementing Select}

Most practical select implementations chunk the bit vector into
\defn{basic blocks}. They also maintain one or more levels of indexes,
called \defn{superblocks}, which identify the basic block containing
the $j$th bit, for some predetermined values of $j$.

The select operation proceeds in two phases.  In the first phase
\textsc{select}$(j)$ does a lookup in the superblocks and finds the
basic block that contains the $j$th one.  In the second phase the
algorithm performs a select operation on the basic block. If the basic
block is larger than a machine word, then the algorithm uses
\textsc{popcount} sequentially to find the machine word that contains
the $j$th bit of the vector.  It then performs a select on that machine
word to find the exact location of the $j$th bit in the vector.

Implementing select on a machine word turns out to be surprisingly
complex, when only standard operations such as shift, mask, multiply,
addition, etc.\ are available.  Therefore, the select operation on
machine words represents a significant component of the time required
to perform select, even on large bit vectors.


For machine words, the select operation has been harder to implement than the
rank operation.
The rank operation on a machine word can be implemented using a dedicated
machine instruction, \textsc{popcount}, which counts the number of 1s in a
machine word.
In contrast, until recently, the best known implementations of select on
machine words used $50$--$80$ machine instructions~\cite{ZhouAnKa13}.  Most
prior implementations use broadword programming~\cite{Vigna08,NavarroPr12},
\textsc{sse4} instructions~\cite{GogPe14} or a combination of broadword
programming and the \textsc{popcount} instruction~\cite{ZhouAnKa13}.

Recently, Pandey et al.~\cite{PandeyBeJo17} gave a new implementation
of machine-word select, which we call \sbu. \sbu uses only four x86 machine
instructions, two of which were introduced in Intel's Haswell CPUs in 2013.
One instruction, \textsc{pdep}, deposits bits from one operand in locations
specified by the bits of the other operand.
The second instruction, \textsc{tzcnt}, returns the number of trailing zeros in
its argument.
Although these two instructions did not exist a few years ago,
they are widely available today.

\subsection*{ Results}

\begin{itemize}

\item We perform an experimental study of how \sbu, the new
  implementation of the select operation on 64-bit words proposed by
  Pandey et al.~\cite{PandeyBeJo17}, affects select on large bit
  vectors.

\item We run two different benchmarks.  First we compare \sbu to
  \cmu~\cite{ZhouAnKa13} and \sdsl~\cite{GogPe14} on machine-word
  select.

\item Second, we replace the machine-word select implementations in
  \cmu and \sdsl with \sbu. We then evaluate the performance gain for
  select on large bit vectors.

\item We show the performance gain as a function of the bit-vector
  size.

\item Our experiments show that \sbu runs $2\times$---$4\times$
  faster than the state-of-the-art implementation~\cite{ZhouAnKa13} on
  64-bit machine words.

\item When we replace the machine-word select implementation  with
  \sbu  on large bit vectors we see a performance gain
  of $20\%$ to $68\%$.

\item We found that the new select not only reduced the number
  of instructions required for each bit-vector select, but also
  improved CPU instruction cache performance and memory-access
  parallelism.

\end{itemize}

\section{ How Select is Being Implemented on Large Bit Vectors}

This section reviews the general bit-vector select implementations in
\cmu~\cite{ZhouAnKa13} and \sdsl~\cite{GogPe14} and explains the role
that machine-word select performs in these algorithms. We then discuss
the differences in the implementations of \cmu and \sdsl.

Both \cmu and \sdsl use a position-based sampling approach to
implement bit-vector select. In position-based sampling, the
bit-vector maintains a side table $S[i]$, where $S[i]=\select(ki)$ for
some constant $k$.  The algorithm thus knows that, for any $y$,
$S[\lfloor y/k\rfloor] \leq \select(y) < S[\lfloor y/k\rfloor + 1]$.
It may then use additional auxiliary data structures to further narrow
down the location of the $y$th one to a single machine word.  After
finding the target machine word, it performs a machine-word select to
find the exact location of the $y$th one in the bit vector.

\subsection{ Select in \cmu}

\cmu groups bits into 512-bit \textit{basic blocks}, groups each group
of 4 basic blocks into a \textit{lower block}, and groups $2^{21}$
lower blocks into an \textit{upper block}.  It maintains a table $L_0$
of the rank of the first bit of each upper block.  A select operation
begins by searching in $L_0$ for the upper block containing the target
bit.  The paper says this is done using binary search; the
implementation performs a linear scan.  For each upper block, \cmu
maintains a table $S$ as described above, which it uses to find a
lower block $B$ near the target bit.  \cmu then consults a table $L_1$
of the rank of the first bit of each lower block.  It uses $L_1$ to
scan forward from $B$ until it finds the lower block $B'$ that
actually contains the target bit.  It then consults a table $L_3$ of
the popcount of each basic block to scan through the basic blocks of
$B'$ until it finds the basic block that contains the target bit.  It
then performs a linear scan of the words in the basic block, using
\textsc{popcount} to find the machine word that contains the target bit.
Finally, it performs a machine-word select to find the precise
location of the target bit.

\cmu optimizes cache performance by using 512-bit basic blocks (which
occupy a single cache line), and by storing $L_1$ and $L_2$ in a
packed, interleaved format so that accessing $L_2$ does not incur any
additional cache miss after accessing the corresponding entry in
$L_1$.  See the paper for details~\cite{ZhouAnKa13}.

\subsection{ Select in \sdsl}

\sdsl divides the bit vector in superblocks by storing position of
every $4096$-th one in a table $S$ as described above. If the size of
a superblock (i.e., difference in the index of $k4096$-th one and
$(k+1)4096$-th one, for some $k$) is larger than or equal to
$log^{4}{n}$, where $n$ is the number of bits in the bit vector, then
it is called \emph{long} and \sdsl maintains another level of
positions by storing the positions of all $4096$ ones. If the size of
a superblock is smaller than $log^{4}{n}$ then it is called
\emph{short}, and \sdsl maintains another level of positions by
storing the positions of every $64$-th one.

 To perform $\select(y)$, the algorithm first performs a lookup in superblocks.
 If the $y$th bit occurs in a long superblock then it can answer by a look up
 in the next level. Otherwise, if the $y$th bit occurs in a short superblock,
 it looks up in the next level to determine the nearest machine word.  The
 algorithm then sequentially performs \textsc{popcount} to determine the target
 64-bit word in which the $y$th bit occurs. It then performs a select on the
 target 64-bit word.

\begin{figure}[t]
  \centering
  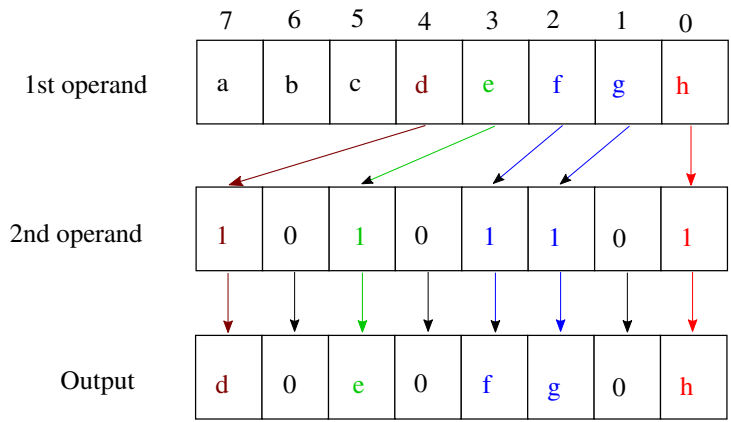
  \caption{A simple example of pdep instruction~\cite{HilewitzLe06}. The 2nd operand acts as a mask
  to deposit bits from the 1st operand.}
  \label{fig:pdep}
\end{figure}

 \begin{algorithm}[t]
   \begin{algorithmic}[1]
     \Function{\sbu}{$x$, $j$}
     \State $i \gets \Call{ShiftLeft}{1, j} $
     \State $p \gets \Call{pdep}{i, x}$
     \State \Return $\Call{tzcnt}{p}$
   \EndFunction
 \end{algorithmic}
 \caption{Algorithm for determining the position of the $j$th 1 in a
   machine word.}
 \label{alg:bitselect}
 \end{algorithm}

\section{ \sbu}

This section reviews Pandey, et al.'s \sbu implementation, shown in
\Cref{alg:bitselect}.  To select the $j$th bit from a 64-bit word $x$,
the algorithm first loads the constant 1 into a register, shifts it
left by $j$, then invokes \textsc{pdep} and \textsc{tzcnt}.

We now explain the \textsc{pdep} and \textsc{tzcnt} instructions,
both of which were introduced in Intel's Haswell line of CPUs.


\textsc{pdep} deposits bits from one operand in locations specified by
the bits of the other operand, as illustrated in \Cref{fig:pdep}.  If
$p=\textsc{pdep}(v,x)$, then the $i$th bit of $p$ is given by
\[
p_i = 
\begin{cases}
  v_j & \text{ if $x_i$ is the $j$th 1 in $x$,} \\
  0   & \text{ otherwise. } \\
\end{cases}
\]

\textsc{tzcnt} returns the number of trailing zeros in its
argument. If $B$ is a 12-bit vector such that
$B[0,11]=$\texttt{110010100000} then $\textsc{tzcnt}(B)=5$.


The \sbu algorithm works because performing $\textsc{pdep}(2^j,x)$
produces a 64-bit integer $p$ with a single bit set---in the same
position as $x$'s $j$th bit.  \textsc{tzcnt} then finds the position
of this bit by counting the number of trailing zeros in $p$.


\section{ Evaluation}

In this section we evaluate \sbu, the new machine-word select implementation
proposed by Pandey et al.~\cite{PandeyBeJo17}.  We compare \sbu against
machine-word select implementations in Zhou at al's \cmu~\cite{rankselect}
(which is an optimized implementation of combined sampling~\cite{NavarroPr12})
and Gog et al.'s \sdsl~\cite{sdsl} (which is an implementation of Clark et
al.~\cite{Clark98} and broadword programming).  We then measure the impact
of \sbu on the performance of the overall bit-vector select algorithms
in these implementations.

We address the following questions about the performance of \sbu:
\begin{itemize}
  \item How does \sbu compare to machine-word select implementations in \cmu
    and \sdsl on 64-bit words in cache?
  \item How does \sbu compare to machine-word select implementations in \cmu
    and \sdsl on 64-bit words not in cache?
  \item How does \sbu affects the performance of bit-vector select in
    \cmu and \sdsl.
\end{itemize}

\subsection{ Experimental Setup}

To answer the above questions, we perform three different benchmarks. First we
evaluate the performance of \sbu and machine-word select implementations in
\cmu and \sdsl on a single 64-bit machine word.
Second we evaluate the performance of \sbu and machine-word select
implementations in \cmu and \sdsl on 64-bit machine words drawn randomly from
a bit vector.
Third we replace the machine-word select implementations in \cmu and \sdsl with
\sbu and evaluate the performance for select operations on large bit vectors.

For our experiments we used the same benchmark suite that is used in Zhou et
al.'s paper~\cite{rankselect}. 
To evaluate \sdsl on the same benchmark, we took the bit vector select
implementation from the \sdsl library~\cite{sdsl} and added it to the benchmark
suite. 
Our version of the benchmarking suite is available at
\url{https://github.com/splatlab/rankselect}.

We performed microbenchmarks to measure how fast \sbu runs on machine
words compared to the machine-word select implementations in \cmu and
\sdsl. In the first benchmark we perform 10M select operations on a
single machine word with random rank values and report the total
time. Because we perform select on a single machine word there are
essentially no cache misses during the experiment.

For the second benchmark, we measure the time required to perform 10M
machine-word select on a word chosen randomly from a bit vector.
Thus, for large bit vectors, the select is likely to incur a cache
miss loading the input word.  This benchmark measures the performance
difference that different machine-word select implementations yield in
a realistic workload of essentially random selects out of a bit
vector.

For the third benchmark, we perform 10M bit-vector selects.  This
benchmark measures the total time to perform the select, including
traversing data structures, performing a machine-word select, and
cache misses.

We perform these experiments with bit vector sizes from $2^{24}$ to
$2^{34}$. For each of the benchmarks, we perform experiments with
three different densities of the bit vector, $90$\%, $50$\%, and
$10$\%.  We perform 10 iterations of each experiment and report the
average time (in nanoseconds) to perform a single operation.

We have also collected some hardware stats in order to analyze our
evaluation results. We have used two tools, \emph{perf} and
\emph{intel VTune} to collect stats.  We used \emph{perf} to collect
the number of instruction-cache misses and the number of
instructions. We used \emph{intel VTune} to collect the average number
of in-flight memory read requests.

All experiments were performed on an Intel(R) Core(TM) i7-6700HQ CPU
(@ 2.60GHz with 4 cores and 6MB LLC) running Ubuntu x86\_64
4.4.0-59-generic.

\subsection{ In-register machine-word select}

\begin{table}
  \begin{centering}
    \begin{tabular}{cS[table-number-alignment = center,
                      table-figures-decimal = 1,
                      table-figures-integer= 2]
                    rrr}
    \hline
    \textbf{Implementation} & \textbf{Time (ns)} \\
    \hline
    \hline
    \cmu & 8.11 \\[.05in] 
    \sdsl & 4.23 \\[.05in] 
    \sbu & 2.07 \\[.05in] 
    \hline
  \end{tabular}
  \caption{ Time to perform a machine-word select on an in-cache
    64-bit machine word by \cmu, \sdsl, and \sbu.  }
    \label{tab:word-time}
  \end{centering}
\end{table}

\Cref{tab:word-time} shows the time taken to perform select on
in-cache 64-bit words.  \sbu is $4\times$ faster than the \cmu
machine-word select and $2\times$ faster than the \sdsl machine-word
select.

\subsection{ Machine-word select}

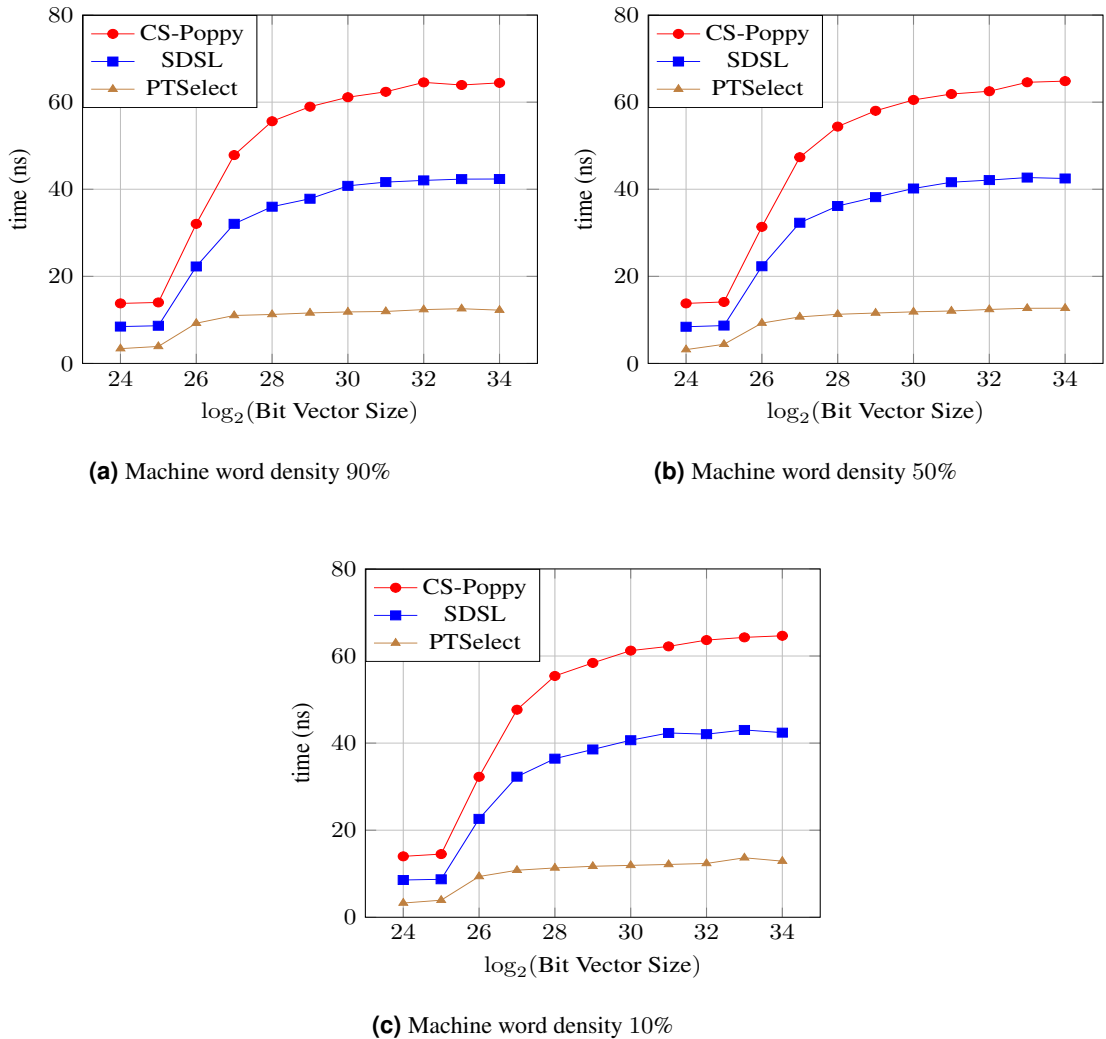
\begin{figure*}[th]
  \captionsetup[subfigure]{justification=centering}
  {\centering
    \begin{subfigure}[t]{.45\textwidth}
      \centering
      \begin{tikzpicture}[yscale=0.85, xscale=0.95]
        \begin{axis}[
            width=\linewidth,
            scale only axis,
            xlabel={$\log_{2}(\text{Bit Vector Size})$},
            ylabel={time (ns)}, 
            xtick={24,26,28,30,32,34},
            ymin=0, ymax=80,
            grid=major,
            legend style={at={(0, 1)},anchor=north west},
            legend entries={\cmu, \sdsl, \sbu},
          ] 
          \addplot[color=red,mark=oplus*]       table {CMUword90.data};
          \addplot[color=blue, mark=square*]    table {SDSLword90.data};
          \addplot[color=brown, mark=triangle*] table {SBUword90.data};
        \end{axis}
      \end{tikzpicture}
      \label{fig:word_select_90}
      \caption{Machine word density $90$\%}
    \end{subfigure}%
    \hspace{3em}
    \begin{subfigure}[t]{.45\textwidth}
      \centering
      \begin{tikzpicture}[yscale=0.85, xscale=0.95]
        \begin{axis}[
            width=\linewidth,
            scale only axis,
            xlabel={$\log_{2}(\text{Bit Vector Size})$},
            ylabel={time (ns)}, 
            xtick={24,26,28,30,32,34},
            ymin=0, ymax=80,
            grid=major,
            legend style={at={(0, 1)},anchor=north west},
            legend entries={\cmu, \sdsl, \sbu},
          ] 
          \addplot[color=red,mark=oplus*]       table {CMUword50.data};
          \addplot[color=blue, mark=square*]    table {SDSLword50.data};
          \addplot[color=brown, mark=triangle*] table {SBUword50.data};
        \end{axis}
      \end{tikzpicture}
      \label{fig:word_select_50}
      \caption{Machine word density $50$\%}
    \end{subfigure}%
    \vspace{2em}
    \begin{subfigure}[t]{.45\textwidth}
      \centering
      \begin{tikzpicture}[yscale=0.85, xscale=0.95]
        \begin{axis}[
            width=\linewidth,
            scale only axis,
            xlabel={$\log_{2}(\text{Bit Vector Size})$},
            ylabel={time (ns)}, 
            xtick={24,26,28,30,32,34},
            ymin=0, ymax=80,
            grid=major,
            legend style={at={(0, 1)},anchor=north west},
            legend entries={\cmu, \sdsl, \sbu},
          ] 
          \addplot[color=red,mark=oplus*]       table {CMUword10.data};
          \addplot[color=blue, mark=square*]    table {SDSLword10.data};
          \addplot[color=brown, mark=triangle*] table {SBUword10.data};
        \end{axis}
      \end{tikzpicture}
      \label{fig:word_select_10}
      \caption{Machine word density $10$\%}
    \end{subfigure}

  }
  \caption{
    Performance of select operation on machine words with density $90$\%,
    $50$\%, and $10$\%. (Lower is better.)
  }
  \label{graph:word_select}
\end{figure*}

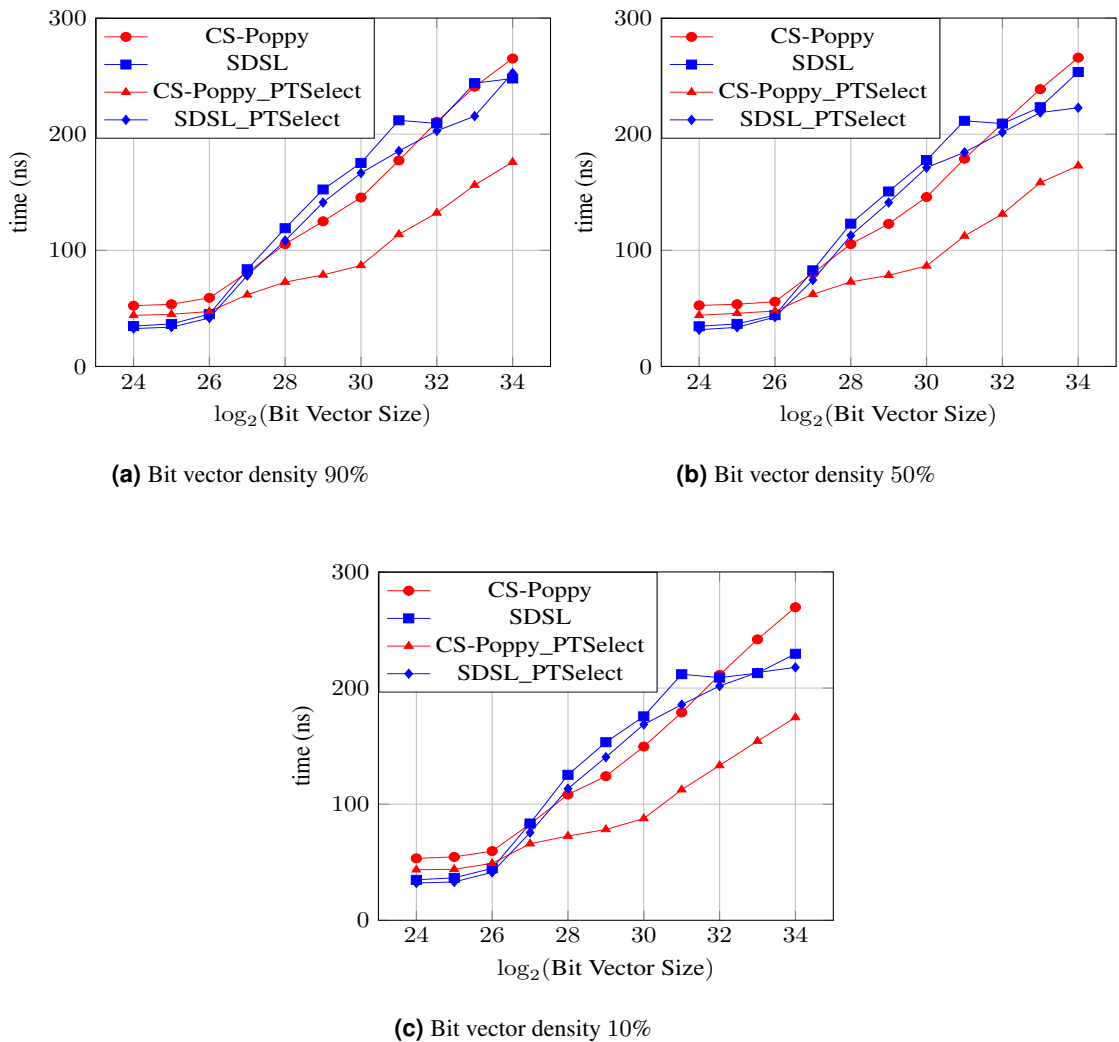
\begin{figure*}[th]
  \captionsetup[subfigure]{justification=centering}
  {\centering
    \begin{subfigure}[t]{.45\textwidth}
      \centering
      \begin{tikzpicture}[yscale=0.85, xscale=0.95]
        \begin{axis}[
            width=\linewidth,
            scale only axis,
            xlabel={$\log_{2}(\text{Bit Vector Size})$},
            ylabel={time (ns)}, 
            xtick={24,26,28,30,32,34},
            ymin=0, ymax=300,
            grid=major,
            legend style={at={(0, 1)},anchor=north west},
            legend entries={\cmu, \sdsl, \cmusbu, \sdslsbu},
          ] 
          \addplot[color=red,mark=oplus*]       table {CMU90.data};
          \addplot[color=blue, mark=square*]    table {SDSL90.data};
          \addplot[color=red, mark=triangle*] table {CMU_SBU90.data};
          \addplot[color=blue, mark=diamond*]  table {SDSL_SBU90.data};
        \end{axis}
      \end{tikzpicture}
      \label{fig:bitvector_select_90}
      \caption{Bit vector density $90$\%}
    \end{subfigure}%
    \hspace{3em}
    \begin{subfigure}[t]{.45\textwidth}
      \centering
      \begin{tikzpicture}[yscale=0.85, xscale=0.95]
        \begin{axis}[
            width=\linewidth,
            scale only axis,
            xlabel={$\log_{2}(\text{Bit Vector Size})$},
            ylabel={time (ns)}, 
            xtick={24,26,28,30,32,34},
            ymin=0, ymax=300,
            grid=major,
            legend style={at={(0, 1)},anchor=north west},
            legend entries={\cmu, \sdsl, \cmusbu, \sdslsbu},
          ] 
          \addplot[color=red,mark=oplus*]       table {CMU50.data};
          \addplot[color=blue, mark=square*]    table {SDSL50.data};
          \addplot[color=red, mark=triangle*] table {CMU_SBU50.data};
          \addplot[color=blue, mark=diamond*]  table {SDSL_SBU50.data};
        \end{axis}
      \end{tikzpicture}
      \label{fig:bitvector_select_50}
      \caption{Bit vector density $50$\%}
    \end{subfigure}%
    \vspace{2em}
    \begin{subfigure}[t]{.45\textwidth}
      \centering
      \begin{tikzpicture}[yscale=0.85, xscale=0.95]
        \begin{axis}[
            width=\linewidth,
            scale only axis,
            xlabel={$\log_{2}(\text{Bit Vector Size})$},
            ylabel={time (ns)}, 
            xtick={24,26,28,30,32,34},
            ymin=0, ymax=300,
            grid=major,
            legend style={at={(0, 1)},anchor=north west},
            legend entries={\cmu, \sdsl, \cmusbu, \sdslsbu},
          ] 
          \addplot[color=red,mark=oplus*]       table {CMU10.data};
          \addplot[color=blue, mark=square*]    table {SDSL10.data};
          \addplot[color=red, mark=triangle*] table {CMU_SBU10.data};
          \addplot[color=blue, mark=diamond*]  table {SDSL_SBU10.data};
        \end{axis}
      \end{tikzpicture}
      \label{fig:bitvector_select_10}
      \caption{Bit vector density $10$\%}
    \end{subfigure}

  }
  \caption{
    Performance of select operation on large bit vector with densities $90$\%,
    $50$\%, and $10$\%. (Lower is better.)
  }
  \label{graph:bitvector_select}
\end{figure*}

\begin{table}
  \begin{centering}
    \begin{tabular}{cS[table-number-alignment = center,
                      table-figures-decimal = 1,
                      table-figures-integer= 2]
                    rrr}
    \hline
    \textbf{Implementation} & \textbf{Instruction count} \\
    \hline
    \hline
    \cmu & 80 \\[.05in] 
    \sdsl & 49 \\[.05in]
    \sbu & 12 \\[.05in]
    \hline
  \end{tabular}
  \caption{ The number of instructions in each implementation of
    machine-word select.  These counts include function prologues and
    epilogues and code for loading the arguments from bit-vector, in
    addition to the core instructions for performing the machine-word
    select. }
    \label{tab:implementation-sizes}
  \end{centering}
\end{table}

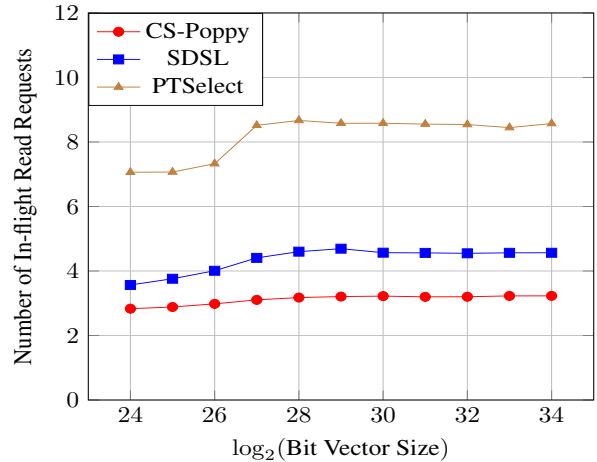
\begin{figure*}[th]
      \centering
       \begin{tikzpicture}[yscale=0.85, xscale=0.95]
        \begin{axis}[
            width=\linewidth/2,
            scale only axis,
            xlabel={$\log_{2}(\text{Bit Vector Size})$},
            ylabel={$\text{Number of In-flight Read Requests}$}, 
            xtick={24,26,28,30,32,34},
            ymin=0,
            ymax=12,
            grid=major,
            legend style={at={(0, 1)},anchor=north west},
            legend entries={\cmu, \sdsl, \sbu},
          ]
          \addplot[color=red,mark=oplus*]       table {CMU_MLP50.data};
          \addplot[color=blue, mark=square*]    table {SDSL_MLP50.data};
          \addplot[color=brown, mark=triangle*] table {SBU_MLP50.data};
        \end{axis}
      \end{tikzpicture}
      \caption{
        \label{fig:mlp}
        The average number of in-flight read requests per cycle when performing
        machine-word select operations on increasing bit vector sizes for \cmu,
        \sdsl, and \sbu.       
      }
\end{figure*}

\Cref{graph:word_select} shows the average time taken per select
operation by \cmu, \sdsl, and \sbu for select operations on machine
words chosen randomly from a bit vector.

\sbu is $2.2\times$--$6\times$ faster than \cmu and \sdsl. \cmu
machine-word select is the slowest for any bit vector size.
For larger bit vectors, \sbu is $5\times$ faster than \cmu and
$3\times$ faster than \sdsl.

We explain these results as follows.  When the bit vector is small,
cache misses are relatively rare, so the cost of each select is
dominated by the computational costs, which are shown in
\Cref{tab:word-time}.  As the bit vector grows, the time to perform a
machine-word select becomes dominated by the time required to load the
queried machine word into cache.

However, this presents one puzzle: if the time to perform a
machine-word select within a large bit vector is dominated by the time
to resolve a cache miss, why is \sbu faster than \sdsl, which is in
turn faster than \cmu?  Shouldn't they all take roughly the same
amount of time, i.e. the time required to serve the cache miss?

The reason they do not all take the same amount of time is
\textit{memory parallelism}.  Modern CPUs can issue multiple memory
reads in parallel, increasing throughput.  \Cref{fig:mlp} shows the
average level of memory parallelism observed for each algorithm during
our machine-word select benchmark.  \sbu has almost 3$\times$ the
parallelism of \cmu, and about 2$\times$ the parallelism of \sdsl.
This largely explains the performance difference between the different
machine-word select algorithms seen for large bit vectors in
\Cref{graph:word_select}.

And why do the different machine-word select implementations enable
different levels of memory parallelism?  Modern CPUs maintain a
relatively large window of outstanding instructions, any of which can
be issued as soon as its inputs are available.  Each iteration of our
benchmark loop is independent of previous iterations, so the CPU can
potentially execute two or more iterations in parallel, if they fit
within its window of outstanding instructions.  Thus, a shorter
implementation of select will enable more iterations to fit in the
CPU's instruction window, enabling greater parallelism.  Each
iteration will induce roughly one cache miss so, as a result, a
shorter select implementation will enable greater memory parallelism.

This is exactly what we see in the benchmark.
\Cref{tab:implementation-sizes} shows the number of instructions in
each select implementation.  All the implementations consist of
straight-line code with no branches.  \cmu consists of 80 instructions
and averages 3 concurrent memory accesses, suggesting an instruction
window of about 240 instructions.  \sdsl consists of 49 instructions
and averages 4 concurrent memory accesses, suggesting an instruction
window of about 200 instructions.  \sbu contains only 12 instructions
and averages 8 concurrent memory accesses, which would suggest a
window size of only about 100 instructions.  We suspect that \sbu is
actually being bottlenecked by the number of outstanding memory
requests supported by the CPU.


\subsection{ Bit-vector select}


\Cref{graph:bitvector_select} shows the average time taken per
bit-vector select by \cmu, \sdsl, \cmu with \sbu and \sdsl with \sbu.

Both \cmu and \sdsl are faster when we replace their default
machine-word select implementation with \sbu. \cmu with \sbu performs
$20$\%--$68$\% faster than traditional \cmu. \sdsl with \sbu performs
$2$\%--$15$\% faster than traditional \cmu.

\section{ Conclusion}

This paper shows that having an efficient implementation of machine-word select
leads to a faster bit-vector select.
%
In our evaluation we found that \sbu is the fastest machine-word select
implementation.
\cmu with \sbu is the fastest bit-vector select for large bit vectors and \sdsl
with \sbu is the fastest bit-vector select for small bit vectors.

Even though the machine-word select is invoked only once during a
bit-vector select (irrespective of the size of the bit vector) the
performance gain by using \sbu increases with increasing bit-vector
sizes.  This is because the small size of \sbu enables the CPU to
execute more selects in parallel, increasing memory-level parallelism,
improving performance for bit vectors that do not fit in cache.
%

We believe that there may be additional opportunities to optimize
succinct data structure implementations using x86 bit-manipulation
instructions.  Furthermore, by making succinct data structures that
match the performance of conventional data structures, we can bring
succinct data structures from theory into practice.

\subparagraph*{Acknowledgments.}

We gratefully acknowledge David Anderson for providing the
benchmarking suite for our comparisons from Zhou et al.'s \cmu.  We
thank Mike Ferdman for suggesting that memory-level parallelism might
explain the performance we observed in our machine-word select
benchmark.

We gratefully 
acknowledge support from 
NSF grants
IIS-1247726, 
IIS-1251137, 
CNS-1408695, 
CCF-1439084, and  
CCF-1617618, 
and from Sandia National Laboratories.




\newpage 
\balance
\bibliography{allpapers,bibliography,BFJ-bigdata-papers}



\end{document}